\documentclass[5p]{elsarticle}

\usepackage[utf8]{inputenc}
\usepackage[T1]{fontenc}

\usepackage{amsmath}
\usepackage{amssymb}
\usepackage{bm}

\usepackage[section]{placeins}
\usepackage{stfloats}
\usepackage{xcolor}

\usepackage{float}
\usepackage{graphicx}
\usepackage{epstopdf}
\usepackage{subcaption}
\usepackage{pgfplots}
\pgfplotsset{compat=1.14}

\usepackage{soul}

\usepackage{dcolumn}

\usepackage{hyperref}
 \hypersetup{
 colorlinks = true,
 linkcolor = black,
 anchorcolor =red,
 citecolor = blue,
 filecolor = red,
 urlcolor = red, 
 pdfauthor=author
 }
 
\usepackage{cleveref}
\usepackage[abbreviations, automake]{glossaries-extra} 
 \makeglossaries
 \newabbreviation{bgk}{BGK}{Bhatnagar, Gross and Krook}
 \newabbreviation{cfd}{CFD}{Computational Fluid Dynamics}
 \newabbreviation{cfl}{CFL}{Courant–Friedrichs–Lewy}
 \newabbreviation{dg}{DG}{Discontinuous Galerkin}
 \newabbreviation{dns}{DNS}{Direct Numerical Simulation}
 \newabbreviation{edf}{EDF}{Equilibrium Distribution Function}
 \newabbreviation{fda}{FDA}{Food and Drug Administration}
 \newabbreviation{fdm}{FDM}{Finite Difference Method}
 \newabbreviation{fem}{FEM}{Finite Element Method}
 \newabbreviation{fvm}{FVM}{Finite Volume Method}
 \newabbreviation{lbm}{LBM}{Lattice Boltzmann Method}
 \newabbreviation{les}{LES}{Large-Eddy Simulation}
 \newabbreviation{mrt}{MRT}{Multiple Relaxation Times}
 \newabbreviation{ns}{NS}{Navier-Stokes}
 \newabbreviation{piv}{PIV}{Particle Image Velocimetry}
 \newabbreviation{rans}{RANS}{Reynolds-Averaged Navier-Stokes}
 \newabbreviation{urans}{URANS}{Unsteady Reynolds-Averaged Navier-Stokes}
 \newabbreviation{wale}{WALE}{Wall-Adapting Local Eddy-viscosity}
 \newabbreviation{srt}{SRT}{single relaxation time}
 \usepackage{easyReview}
\usepackage{lineno}
\journal{Computer Methods and Programs in Biomedicine}
\begin{document}
\begin{frontmatter}
 \title{Simulation of the FDA Nozzle Benchmark: A Lattice Boltzmann Study}
 \author[OvGU]{Feng Huang}
 \author[UGE]{Romain No\"el}
 \author[OvGU,STIM]{Philipp Berg}
 \author[OvGU,ETHZ]{Seyed Ali Hosseini\corref{mycorrespondingauthor}}
 \cortext[mycorrespondingauthor]{Corresponding author}
 \ead{seyed.hosseini@ovgu.de}
 
 \address[OvGU]{Laboratory of Fluid Dynamics and Technical Flows, University of Magdeburg ``Otto von Guericke'', D-39106 Magdeburg, Germany}
 \address[ETHZ]{Department of Mechanical and Process Engineering, ETH Z\"urich, 8092 Z\"urich, Switzerland}
 \address[UGE]{Univ. Gustave Eiffel, Inria, Cosys/SII, I4S, F-44344 Bouguenais, France}
 \address[STIM]{Research Campus STIMULATE, University of Magdeburg ``Otto von Guericke'', D-39106, Magdeburg, Germany}

 \begin{abstract}
\emph{Background and objective}: Contrary to flows in small intracranial vessels, many blood flow configurations such as those found in aortic vessels and aneurysms involve larger Reynolds numbers and, therefore, transitional or turbulent conditions. Dealing with such systems require both robust and efficient numerical methods.\\

\noindent\emph{Methods}: We assess here the performance of a lattice Boltzmann solver with full Hermite expansion of the equilibrium and central Hermite moments collision operator at higher Reynolds numbers, especially for under-resolved simulations. To that end the food and drug administration's benchmark nozzle is considered at three different Reynolds numbers covering all regimes: 1) laminar at a Reynolds number of $500$, 2) transitional at a Reynolds number of $3500$, and 3) low-level turbulence at a Reynolds number of $6500$.\\

\noindent\emph{Results}: The lattice Boltzmann results are compared with previously published inter-laboratory experimental data obtained by particle image velocimetry. Our results show good agreement with the experimental measurements throughout the nozzle, demonstrating the good performance of the solver even in under-resolved simulations.\\

\noindent\emph{Conclusion}: In this manner, fast but sufficiently accurate numerical predictions can be achieved for flow configurations of practical interest regarding medical applications.
\end{abstract}
\end{frontmatter}

\section{Introduction}
 \gls{cfd} is known to be a demanding field in terms of computational resources and is sensitive to the numerical scheme employed. Traditional numerical methods used in \gls{cfd} are computationally heavy and rely on relatively fastidious discrete operations, especially when the geometry is complex. The \gls{lbm} is an alternative numerical technique that gained increasing attention, particularly so over the last two decades.
 Its popularity is due to the simplicity of the involved numerical operations, their locality -- ensuring excellent portability to parallel computing architectures -- and the relative ease at which complex geometries can be treated ~\cite{succi2001lattice}. These strengths combined with its ability to model mechano-biological phenomena~\cite{noel2017lattice,Ouared_2008}, make the \gls{lbm} a method of choice for the bio-medical community~\cite{hosseini2021central,jain2020efficacy,Hasert_2014,Malaspinas_2016,mccullough2021efficient,mccullough2020high,kadri2019vivo,khosravanian2021fast,afrouzi2020simulation}.
 
 Nevertheless, computational costs are still often prohibitive in high Reynolds number flows for the classical \gls{lbm}, for instance in aortic aneurysms. 
 That is mainly due to the fact that the standard \gls{lbm} using a collision operator with a single parameter rate relaxation toward a second order expanded \gls{edf}, suffers from shortcomings such as limited stability domain and lack of Galilean invariance~\cite{prasianakis2008lattice,Dellar_2003,hosseini2019stability}.

 Thus, many modifications have been proposed to extend the accuracy and stability of the \gls{lbm}, most notably, through more advanced collision operators, e.g. the entropic formulation~\cite{Ansumali_2002,Karlin_2014}, the \gls{mrt}~\cite{dHumieres_2002}, centered or cascaded~\cite{Geier_2006}, the Cumulant method~\cite{Geier_2015}, or the regularized \gls{lbm}~\cite{Malaspinas_2015}. 
 Furthermore, the inconsistency in the shear components can be readily overcome by using higher order expansions in the \gls{edf}~\cite{Dellar_2003}. 
 The development of more consistent and robust schemes has also opened the door to the so-called under-resolved direct numerical simulations~\cite{hosseini2019stability,geier2017parametrization}. 
 A number of studies have shown that such simulations can match \gls{les} with explicit sub-grid scale closures in terms of accuracy in many configurations, as demonstrated most recently in \cite{geier2021under} for homogeneous isotropic turbulence. 
 In the present work, we will further explore that avenue in the context of biological flows, through a well-established benchmark.

 The need for reliable, reproducible benchmarks confronted with experiments is urging in bio-medical applications where many variables may affect the results: temperature, in-vivo vs. ex-vivo vs. in-vitro, biological species, etc. The \gls{fda}, conscious of this reality, suggested a series of challenging benchmarks. The idealized medical nozzle device benchmark has been used by several laboratories to conduct \gls{piv} measurements.
 Those measurements provided reference standards to test the accuracy, stability and efficacy for numerical solvers or alternative experimental measurement techniques~\cite{hariharan2011multilaboratory}.
 \begin{table*}[th]
 \centering
 \footnotesize
 \setlength{\tabcolsep}{2pt}
 \caption{Summary of numerical studies associated to the \glsentryshort{fda} benchmark nozzle. \label{Table:summaryArticles}}
 \begin{tabular}{||cccccccc||}
 \hline
 Study & Re$=500$ & Re$=2000$ & Re$=3500$ & Re$=5000$ & Re$=6500$ & Numerical method & Sub-grid model\\
 \hline\hline
 Sanchez Abad et al. \cite{sanchez2020simulation} & \checkmark & \checkmark & \checkmark & \checkmark & & Spectral elements (Nek5000) & None\\
 \hline
 Fehn et al. \cite{fehn2019modern} & \checkmark & \checkmark & \checkmark & \checkmark & \checkmark & High-order \glsentryshort{dg} & None\\
 \hline
 Bergersen et al. \cite{bergersen2019fda} & \checkmark & & \checkmark & & \checkmark & \glsentryshort{fem} (Oasis) & None\\
 \hline
 Pewowaruk et al. \cite{pewowaruk2021solution} & \checkmark & & \checkmark & & \checkmark & \glsentryshort{fvm} (Converge v2.4) & None/\glsentryshort{rans}\\
 \hline
 Stewart et al. \cite{stewart2012assessment} & \checkmark & \checkmark & \checkmark & \checkmark & \checkmark & various & various\\ 
 \hline
Zmijanovic et al. \cite{zmijanovic2017numerical} & & & \checkmark & & & \glsentryshort{fvm} (YALES2BIO) & $\sigma$-model\\
 \hline
 Manchester et al. \cite{manchester2020effect} && \checkmark &&&& \glsentryshort{fvm} (OpenFOAM) & \glsentryshort{wale}\\
 \hline
 Bhushan et al. \cite{bhushan2013laminar} & \checkmark && \checkmark && \checkmark & \glsentryshort{fvm} (ANSYS Fluent) & \glsentryshort{urans}\\
 \hline
 Janiga \cite{janiga2014large}& & & & & \checkmark & \glsentryshort{fvm} (ANSYS Fluent) & Smagorinsky\\
 \hline
 Chabannes et al. \cite{chabannes2017high} & \checkmark & \checkmark & \checkmark & & & \glsentryshort{fem} (Feel++) & None\\
 \hline
 White et al. \cite{white2011rotational} & \checkmark & & & & & regularized \glsentryshort{lbm} (Palabos) & None\\
 \hline
 Jain \cite{jain2020efficacy} & & \checkmark & \checkmark & & & raw moments \glsentryshort{mrt}-\glsentryshort{lbm} (Musubi) & None\\
 \hline
 Present work & \checkmark & & \checkmark & & \checkmark & central Hermite \glsentryshort{mrt}-\glsentryshort{lbm} (ALBORZ) & None\\
 \hline\hline
 \end{tabular}
 \vspace{-4mm}
 \end{table*}
 Experimental data sets have been provided by independent laboratories at different Reynolds numbers Re$= 500,\ 2000,\ 3500,\ 5000,$ and $6500$, computed using the diameter of the throat (see later \cref{fig:nozzle_crossSectionView}). These values of Reynolds numbers are typically encountered in medical devices involving flow recirculations, contractions, or expansions. This range of Re covers very different flow regimes going from laminar (for Re = 500 and 2000) to transitional (for Re = 3500 and 5000) to low-level turbulence (for Re = 6500). The corresponding \gls{piv} experiments have been carried out completely independently by three different groups. It must be kept in mind that large inter-laboratory discrepancies have been observed, in particular regarding the jet breakdown location for transitional and turbulent flows. 
 Conversely, a very good agreement was observed for fully laminar flows~\cite{stewart2012assessment}.
 
 Many numerical studies of this benchmark nozzle have been performed over the years, employing various numerical schemes ranging from \gls{fem}~\cite{bergersen2019fda,chabannes2017high} to \gls{fvm}~\cite{zmijanovic2017numerical,janiga2014large}, as well as a few using \gls{lbm}~\cite{jain2020efficacy,white2011rotational}.
 \Cref{Table:summaryArticles} summarizes the published numerical studies of the \gls{fda} benchmark nozzle with the associated Reynolds numbers, numerical methods and turbulence models employed. 
 While many studies performed \gls{dns} for the lower Reynolds numbers, most simulations in the turbulent regime were conducted using turbulence models like \gls{urans}~\cite{pewowaruk2021solution,bhushan2013laminar} or \gls{les}~\cite{manchester2020effect,janiga2014large,delorme2013large}. 
 
 Few studies simulated the \gls{fda} nozzle benchmark using \gls{lbm}; none of those has covered all experimental conditions, more specifically the higher Reynolds numbers. White and Chong~\cite{white2011rotational} performed \gls{lbm} simulations at low Reynolds numbers ($50$ -- $500$) and studied the suitability of different lattice types. More recently, Jain~\cite{jain2020efficacy} studied the \gls{fda} nozzle at Re=2000 and 3500 with a simple \gls{lbm} scheme (raw moments \gls{mrt} collision operator with second order \gls{edf} and D3Q19 lattice) focusing on fully-resolved simulations. Results were shown for the average axial velocity, pressure profiles, and shear stress in different planes.
 In the present study, we consider a wider range of Reynolds number, covering all flow regimes. The computations are done with an advanced \gls{lbm} model relying on a central Hermite collision operator with a full expansion of the discrete equilibrium, implemented in the in-house \gls{lbm} solver ALBORZ~\cite{hosseini2019theoretical,Hosseini_2019a,hosseini2020development,hosseini2020low}. The accuracy of this numerical model is assessed in under-resolved direct numerical simulations at different resolutions, akin to filtered simulations. A comprehensive comparison to experimental data covering all relevant fields is presented.
 
 The article is organized as follows:
 \Cref{sec:numericalMethod} is dedicated to the numerical methods underlying our \gls{lbm} model.
 \Cref{sec:benchmark} details geometries and conditions used. The obtained results are presented in \Cref{sec:results} and discussed in \Cref{sec:discussion}. \Cref{sec:conclusion} brings concluding lights on this study.  

\section{Methods\label{sec:numericalMethod}}
\subsection{Numerical model}
 The flow of interest here, involving a fluid with properties similar to water at room temperature and therefore a sound speed of $c_s\approx 1450~\rm{m/s}$, and a maximum characteristic speed of the order of $10\,\rm{m/s}$, falls into the low-Mach regime, with Ma below $7\times10^{-3}$. 
 As such it can readily be described with a low-Mach approximation to the \gls{ns} equations:
 \begin{equation}
 \partial_t (\rho u_\alpha) + \partial_\beta (\rho u_\alpha u_\beta) + \partial_\beta \mathcal{T}_{\alpha\beta} = 0,
 \end{equation}
 where the stress tensor $\mathcal{T}$ is:
 \begin{equation}
 \mathcal{T}_{\alpha\beta} = p\delta_{\alpha\beta} - \mu\left(\partial_\beta u_\alpha+\partial_\alpha u_\beta - \frac{2}{D}\partial_\gamma u_\gamma\delta_{\alpha\beta}\right) - \eta\partial_\gamma u_\gamma\delta_{\alpha\beta},
 \end{equation}
 {where} $p$, $\mu$, $\eta$ and $D$ are respectively the pressure, dynamic viscosity, bulk viscosity and number of dimensions in space; $\delta_{\alpha\beta}$ denotes the Kronecker delta function.
 
 Low-Mach flows can be dealt with in a variety of ways: using compressible \gls{ns} solvers would inevitably result in rather small time-steps and excessive computational costs due to the large discrepancy between the acoustic and convective modes in terms of propagation speeds; or incompressible \gls{ns} solvers with an infinite sound speed (i.e. $\rm{Ma}=0$) which, while overcoming the stiffness stemming from acoustics, introduce a non-local elliptic equation for the pressure (Poisson equation). 
 A third approach, in-between these two, is a compressible solver with rescaled acoustics (still guaranteeing separation of scales between acoustics and hydrodynamics) with larger effective Mach number allowing for a fully hyperbolic/parabolic system of equations and larger time-steps compared to the fully compressible formulation.
 The now {well-known} and widely used \gls{lbm} targets the latter approximation with a discrete set of evolution equations for discrete probability distribution functions, $f_i$~\cite{kruger2017lattice}:
 \begin{equation}
 f_i(\bm{x}+\bm{c}_i\delta t, t+\delta t) - f_i(\bm{x}, t) = \frac{\delta t}{\bar{\tau}}\left(f_i^{\rm eq}(\rho,\bm{u}) - f_i(\bm{x}, t)\right),
 \label{eq:LBGKE}
 \end{equation}
 where $\bm{c}_i$ are discrete particle velocities, $\bar{\tau}$ the relaxation time, $\delta t$ the time-step size and $f_i^{\rm eq}$ the discrete equilibrium distribution function.
 This system of equations is readily obtained by integrating the discrete-in-phase-space Boltzmann equation with the \gls{bgk} collision operator along its characteristics~\cite{he1997theory,he1997priori}. 
 The discrete-in-phase-space Boltzmann equation is obtained by projecting the continuous space of particle velocities of dimension $D$ onto a set of ortho-normal base functions, e.g, Hermite polynomials, and keeping lower-order contributions needed to recover the correct hydrodynamic limit (i.e. \gls{ns} level dynamics)~\cite{shan2006kinetic}.
 As such the discrete equilibrium function is a truncated Hermite expansion of the Maxwell-Boltzmann equilibrium function~\cite{shan2006kinetic}:
 \begin{equation}\label{eq:equilibrium}
 f_i^{\rm eq}\left(\rho, \bm{u}\right) = \sum_{n=0}^{N} \frac{w_i}{i!\theta_0^{i}} \bm{a}^{\rm eq}_n(\rho,\bm{u}):\mathcal{H}_n(\bm{c}_i),
 \end{equation}
 where {for a third-order quadrature} the stencil reference temperature is $\theta_0=\delta r^2/3\delta t^2$, $\mathcal{H}_n(\bm{c}_i)$ and $\bm{a}^{\rm eq}_n(\rho,\bm{u})$ are the Hermite polynomials and corresponding equilibrium coefficients, while the weights $w_i$ and the set of discrete particle velocities $c_i$ are obtained from a Gauss-Hermite quadrature.
 
 While most of the initial models and simulations relied on second order expansion of the equilibrium distribution, it has since been observed that this leads to Galilean invariance issues in the shear components of the stress tensor. This problem can readily be removed by adding third-order components to the discrete equilibrium~\cite{dellar2014lattice}.
 The same issue exists for normal components of the viscous tensor. However, this problem can only be removed via correction terms~\cite{prasianakis2008lattice,dellar2014lattice,hosseini2020compressibility}.
 The present study targets the incompressible regime where bulk viscosity effects are negligible. 
 Therefore, this correction is not necessary and will not be considered. 
 
 The relaxation time appearing in the evolution equation is tied to the local dynamic viscosity as:
 \begin{equation}
 \bar{\tau} = \frac{\mu}{\rho \theta_0} + \frac{\delta t}{2}.
 \end{equation}
 The original model, as described in \cref{eq:LBGKE} is referred to as the single-relaxation time realization of the \gls{bgk} collision operator, which has a limited range of stability. 
 To enhance the domain of stability and have a more robust scheme allowing for under-resolved simulations we use a multiple relaxation time realization based on central Hermite polynomials. Then, \cref{eq:LBGKE} changes into:
 \begin{equation}
 f_i(\bm{x}+\bm{c}_i\delta t, t+\delta t) - f_i(\bm{x}, t) = \bm{T}^{-1}\bm{S}\bm{T}\left(f_i^{\rm eq}(\rho,\bm{u}) - f_i(\bm{x}, t)\right)
 \end{equation}
 where $\bm{T}$ and $\bm{T}^{-1}$ are the moments transform tensor and its inverse defined as:
 \begin{equation}
 \bm{T}\cdot\bm{f} = \widetilde{\bm{M}}, 
 \end{equation}
 and $\bm{S}$ is the diagonal tensor of relaxation coefficients. 
 The central Hermite moments $\widetilde{\bm{M}}$ are defined as~\cite{hosseini2021central,hosseini2020development}:
 \begin{equation}
 \widetilde{M}_n = \widetilde{a}_n = \sum_i \mathcal{H}_n(\bm{c}_i-\bm{u})f_i.
 \end{equation}
 While relaxation times appearing in $\bm{S}$ can be tuned individually for each moment, here all ghost relaxations are set to one, as this both minimizes computational costs and is very close to the optimal linear stability manifold in the space of ghost relaxation times~\cite{hosseini2020development}. For more details on the transforms and moments readers are referred to~\cite{hosseini2021central,hosseini2020compressibility}.

\subsection{Numerical simulations\label{sec:benchmark}}
\begin{figure*}
 \centering
 \includegraphics[width=0.7\textwidth]{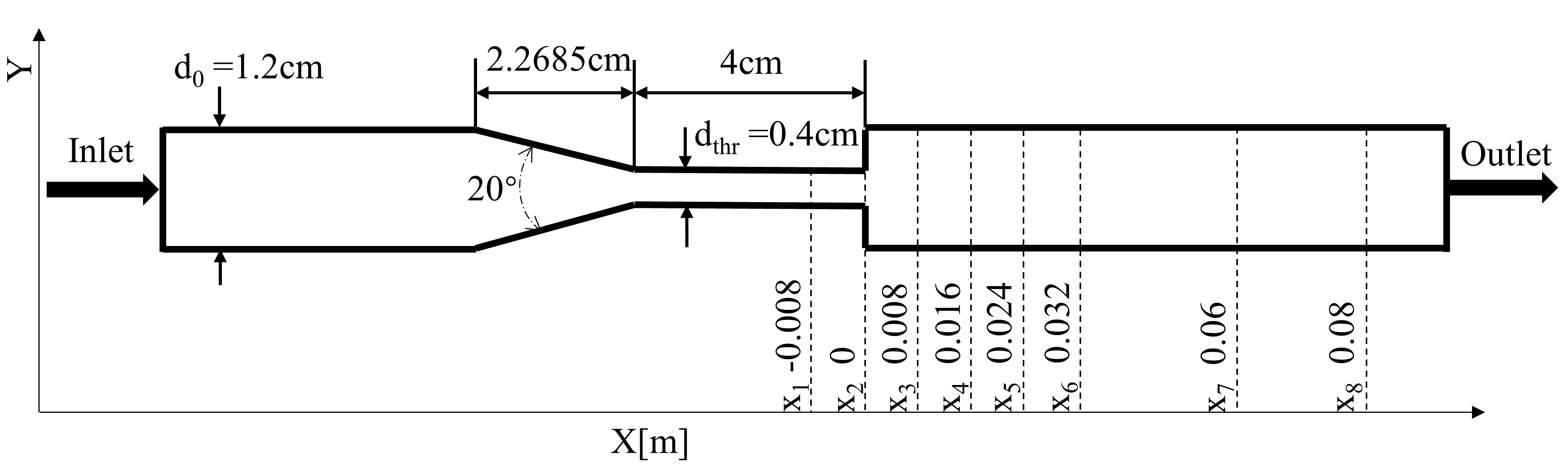}
 \caption{\label{fig:nozzle_crossSectionView}Schematic representation of the \gls{fda} nozzle with all dimensions.}
 \end{figure*}

\paragraph{Cases description}
 A cross-sectional view of the idealized medical device, subject of the \gls{fda} nozzle benchmark is shown in \Cref{fig:nozzle_crossSectionView}.
 The geometry is composed of an axisymmetric nozzle with a convergent section on the flow inlet extremity, a constant throat section in the middle and a sudden expansion section toward the end. 
 To eliminate a possible numerical effect of inlet and outlet boundary conditions, the nozzle has been extended by $0.052\,\rm{m}$ and $0.02\,\rm{m}$ at both ends, respectively. 
 The final length of the nozzle is $0.24\,\rm{m}$. To validate the numerical results, data is collected from both the horizontal center-line and eight distinct planes located at different coordinates along the horizontal $x$-axis. \Cref{fig:nozzle_crossSectionView} shows the locations of these planes. 
 The simulations are run over a total period of time covering $300\ t_c$, where $t_c$ is the flow-through period defined as,
 \begin{equation}
 t_c = \frac{L_c}{u_c},
 \end{equation}
 where the characteristic speed $u_c$ and length $L_c$ are set to the velocity in the throat $u_{\rm thr}$ and to the length of the throat section, respectively. Only data from the second half, i.e. $150-300\ t_c$ is used for the time-averaging process to get average flow fields. The \gls{piv} data used for validation and the 3-D model geometry are available from the \gls{fda} official website\footnote{\protect\url{https://ncihub.org/wiki/FDA_CFD}}.
 
\paragraph{Flow conditions}
Following experimental settings and benchmark requirements, the fluid was considered to be Newtonian, with a density of $1056\,\rm{kg/m^3}$ and a dynamic viscosity of $3.5\,\rm{mPa\,s}$. The Reynolds number is computed as:
 \begin{equation}
 \hbox{Re}= \hbox{Re}_{\rm thr} = \frac{\rho\overline{u}_{\rm thr}d_{\rm thr}}{\mu},
 \end{equation}
 where $d_{\rm thr}$ is the throat diameter and $\overline{u}_{\rm{thr}}$ is defined as:
 \begin{equation}
 \overline{u}_{\rm thr} = {\left(\frac{d_0}{d_{\rm thr}}\right)}^2 \overline{u}_0.
 \end{equation}
 The average inlet velocity $\overline{u}_0$ is obtained dividing the volume flow rate by the inlet surface area $S_0=\pi d_0^2/4$. To cover all flow regimes, the present study considers Re=500, 3500, and 6500. Based on critical pipe Reynolds numbers, the first would fall in the laminar regime, the second in the transitional, and the third in the low-level turbulent regime, respectively.

\paragraph{Simulation details}
The inlet boundary condition for the velocity is prescribed in the form of a laminar parabolic profile:
 \begin{equation}
 u(r) = 2{\overline{u}_0}\left(1-4\frac{r^2}{d_0^2}\right).
 \label{InletVelProfile}
 \end{equation}
 Constant pressure boundary conditions are implemented at the outlet while no-slip boundary conditions are employed along the walls. The no-slip conditions are enforced with the curved treatment of the bounce-back method~\cite{bouzidi2001momentum} while the outflow constant pressure is applied using the non-equilibrium extrapolation approach~\cite{zhao2002non}.
 
 The computational domain is tridimensional ($D=3$) and has a volume of $0.012\times 0.012\times0.24\,\rm{m^3}$. Simulations were conducted with {different grid resolutions}, $\delta r$, listed in \cref{tab:resolution_core}. The respective time-step sizes, $\delta t$, are defined by fixing the maximum predicted convective \gls{cfl} number in the domain:
 \begin{equation}
 \hbox{CFL} = \frac{2\overline{u}_{\rm thr}}{\delta r/\delta t},
 \end{equation}
 in turn leading to:
 \begin{equation}
 \delta t = \frac{\rho d_{\rm thr}\delta r\rm{CFL}}{2\mu\rm{Re}_{\rm thr} }.
 \end{equation}
For the analysis of the turbulent statistics, the velocity is decomposed into a mean and a fluctuating part as:
 \begin{equation}
 \bm{u}(\bm{x},t) = \overline{\bm{u}}(\bm{x}) + \bm{u'}(\bm{x},t),
 \end{equation}
 where the mean part $\overline{\bm{u}}$ is defined by:
 \begin{equation}
 \overline{\bm{u}}(\bm{x}) = \frac{1}{\Delta t}\int_{t_0}^{t_0+\Delta t} \bm{u}(\bm{x},t) dt,
 \end{equation}
 $t_0$ indicates the start of averaging and $\Delta t$ the period over which it is done. Averaging the convective momentum flux term in the \gls{ns} equations using this decomposition, one gets:
 \begin{equation}
 \overline{({{u}}_\alpha + {u'}_\alpha)({{u}}_\beta + {u'}_\beta)} = \overline{{u}_\alpha {u}_\beta} + \overline{{u'}_\alpha {u'}_\beta},
 \end{equation}
The last term of this equation appears in the Reynolds stress tensor, defined as:
 \begin{equation}
 \Pi'_{\alpha\beta} = \rho\overline{{u'}_{\alpha} {u'}_{\beta}},
 \end{equation}
 where $\bm{u'}(\bm{x},t)$ and $\rho$ indicate velocity fluctuations and fluid density, respectively.
 The Reynolds stress tensor is symmetrical and consists of three normal stress and six shear stress components. 
 The normal stress can be calculated as $\rho \overline{u'_\alpha u'_\alpha}$ where the term $\overline{u'_\alpha u'_\alpha}$ indicates the time-averaged product of normal velocity fluctuations. 
 Finally, the viscous shear stress $\sigma$ is calculated, in this study, as below:
 \begin{equation}
     \sigma_{\alpha\beta} = \rho\nu\left(\partial_\beta u_\alpha+\partial_\alpha u_\beta\right).
 \end{equation}
 \begin{table}[htb]
 \centering
 \footnotesize
 \setlength{\tabcolsep}{2pt}
 \caption{Spatial and temporal resolutions of simulations.}
 \begin{tabular}{|cccc|}
 \hline
 & Re & $\delta r$ [$\times10^{-4}\rm{m}$] & $\delta t$ [$\times10^{-6}\rm{s}$]\\
 \hline
 \hline
 R1 & $500$ & $2.3$ & $49.03$ \\
 \hline
 R2 & $500$ & $1.725$ & $36.75$ \\
 \hline
 R3 & $500$ & $1.15$ & $24.5 $\\
 \hline
 \hline
 R1 & $3500$ & $2.3$ & $9.86$  \\
 \hline
 R2 & $3500$ & $1.725$ & $7.38$ \\
 \hline
 R3 & $3500$ & $1.15$ & $4.93$ \\
 \hline
 \hline
 R1 & $6500$ & $2.3$ & $5.31$ \\
 \hline
 R2 & $6500$ & $1.725$ & $3.98$ \\
 \hline
 R3 & $6500$ & $1.15$ & $2.65$ \\
 \hline
 R4 & $6500$ & $0.776$ & $1.7$ \\
 \hline
 \end{tabular}
 \label{tab:resolution_core}
 \end{table}
{Four different resolutions will be considered in the study, from R1 (coarsest) to R4 (finest),} listed in \cref{tab:resolution_core}.
 
\section{Results\label{sec:results}}
Numerical results are compared to experimental data from three independent laboratories, presented in the form of five different data-sets~\cite{hariharan2011multilaboratory}, i.e. three different measurements from the first laboratory, and one each for the second and third laboratories. Throughout this section the three independent sets of measurements from the first laboratory are shown with diamonds, circular, and square markers. Data from the second and third laboratories are represented by asterisks and upward-pointing triangular markers, respectively.
\subsection{Laminar flow, Re = 500}
\begin{figure}[h]
 	\centering
 	\includegraphics[width=0.35\textwidth]{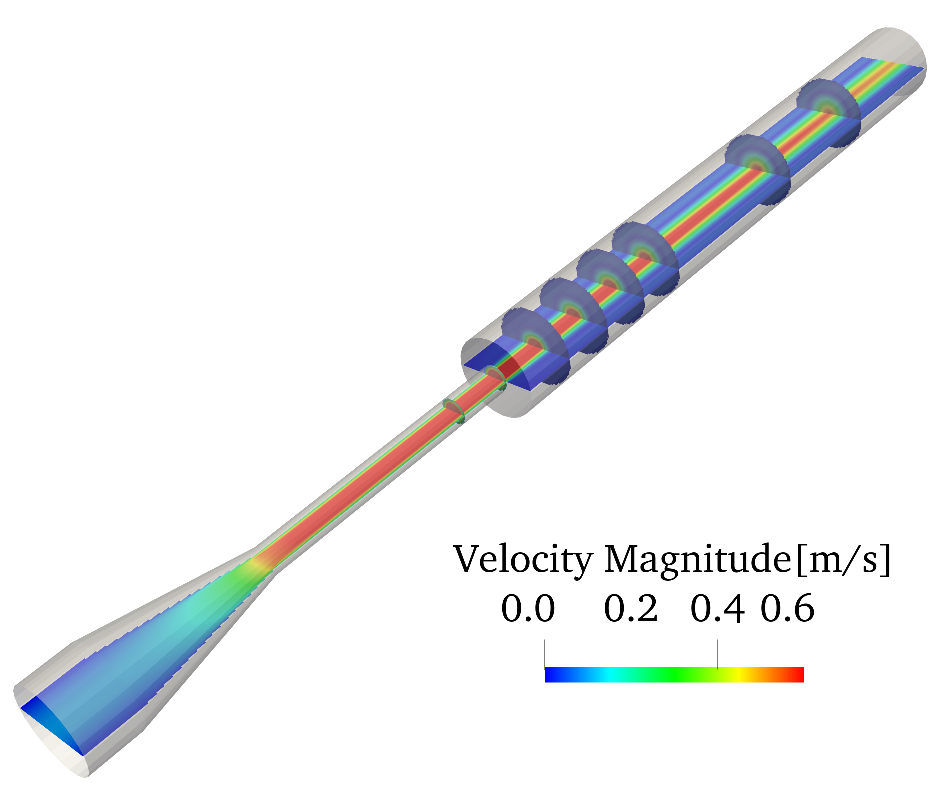}
 	\caption{\label{fig:Re500FlowField} Results for Re=500. Snapshot of the steady velocity field as obtained from R2 simulations at time $t=2.4$~s.}
 \end{figure}

\begin{figure}[htbp]
 \centering
 \includegraphics{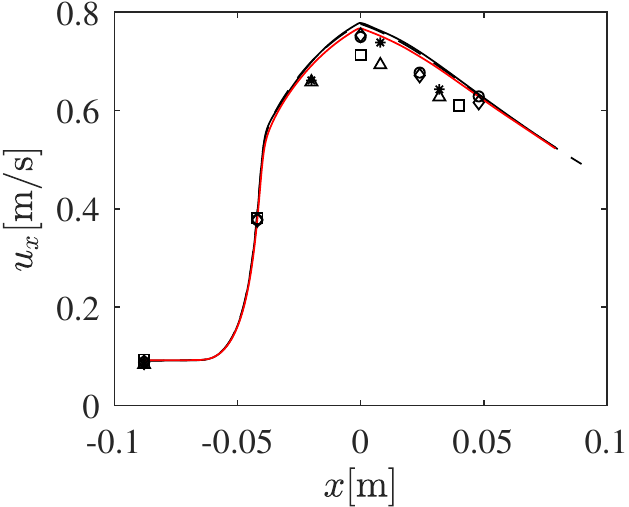}
 \caption{Results for Re = 500. Comparison of mean axial velocity between \gls{lbm} results and \gls{piv} experiments along the center-line of the nozzle. The results from R1, R2, and R3 are indicated by {solid black, dashed black and solid red} lines respectively.}
 \label{fig_Re500pressureAndVelocity}
 \end{figure}

\begin{figure*}[bp]
 \centering
 \includegraphics[width=0.8\textwidth]{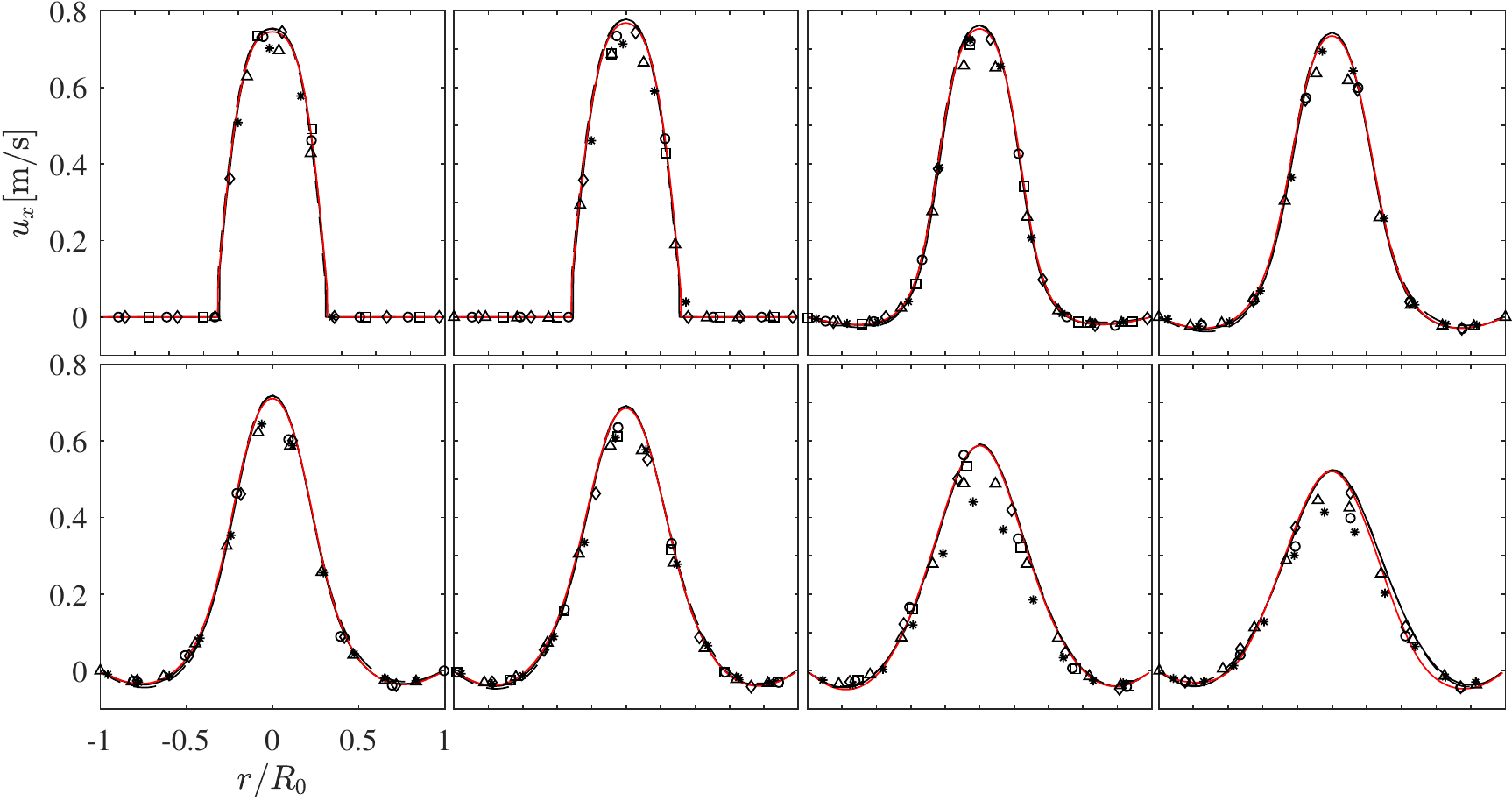}
 \caption{\label{fig:Re500Axial_Vel_RadialDirection} Results for Re = 500. Axial velocity profiles compared with \gls{piv} experiments at selected cross-sections. The radial distance is normalized by the radius of the nozzle, $R_0$. The results from R1, R2, and R3 are indicated by {solid black, dashed black and solid red} lines respectively. Cross-section planes X$_1$ to X$_8$ are represented from top left to bottom right.}
 \end{figure*}

At Re $=500$ (inlet Reynolds number of 167), the flow regime is expected and has been observed to be fully laminar from both \gls{piv} and numerical results, meaning that no jet break-down is observed after the sudden expansion (see \Cref{fig:Re500FlowField}).
\paragraph{Axial velocity}
The results as obtained from R1, R2 and R3 resolutions are compared to experimental data in \Cref{fig_Re500pressureAndVelocity,fig:Re500Axial_Vel_RadialDirection}. First, we look into the time-averaged axial velocity (note that, in this laminar configuration, time-averaged fields are equivalent to the steady fields) along the center-line. All three numerical profiles closely match each other -- lying often on top of each other --, indicating that the chosen grid-sizes allow to correctly resolve the flow features.

It can be noted from the axial velocity plot that the flow is constant upstream of the contraction section ($x < -0.063$~m) and accelerates when flowing into the contraction region, reaching a peak value of 0.76~m/s at the entrance of the expansion section, plane $\rm{X_2}$ (corresponding to $x=0$). The flow starts decelerating with an approximately linear rate in the sudden expansion section $(x > 0)$ due to the increase in cross-section area.
In both \Cref{fig_Re500pressureAndVelocity,fig:Re500Axial_Vel_RadialDirection}, large discrepancies are observed between the different measurements. Stewart et al.~\cite{stewart2012assessment} reported that the experimental data-sets are strongly affected by normalization errors, which could partly explain the observed variations.

\paragraph{Radial profiles of axial velocity on sampling planes}
The profiles of the predicted axial velocity are illustrated at eight cross-sections in \Cref{fig:Re500Axial_Vel_RadialDirection}. A good quantitative agreement is observed between \gls{piv} data and numerical results throughout the nozzle.

However, the peak values of the axial velocity appear to be slightly overestimated after the sudden expansion section. Hariharan et al.\cite{hariharan2011multilaboratory} reported that there is a less than 1\% fluctuation in flow rate during the experiments, which can lead to this difference.  Given that discrepancies between experimental and numerical results are at the same level as measurement uncertainties, it can still be concluded that numerical simulations agree well with the experiments for this case. 

\begin{figure*}[tp]
 \centering
 \includegraphics[width=0.8\textwidth]{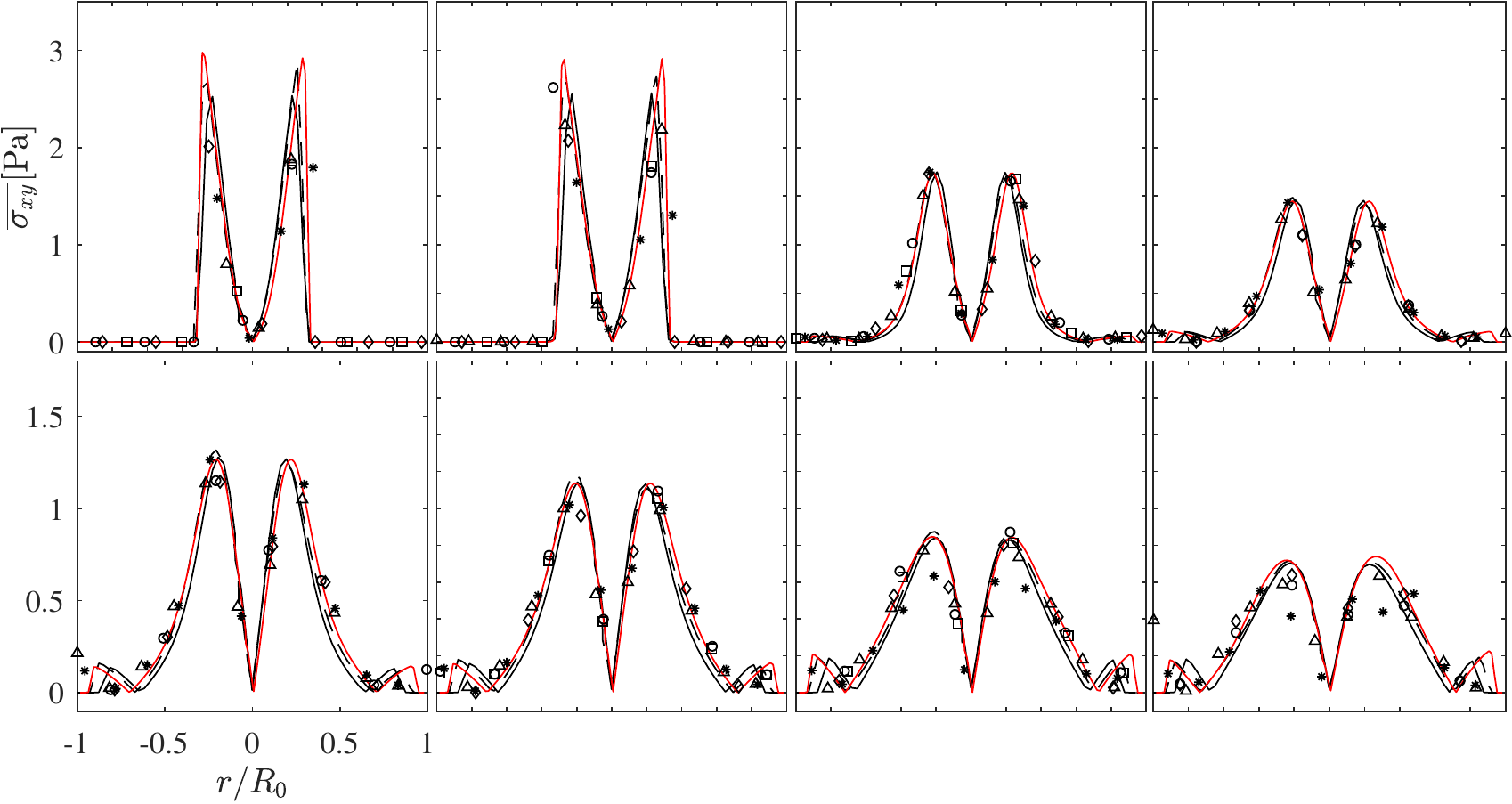}
 \caption{\label{fig:Re500ViscousShearStress} Results for Re = 500. Viscous shear stress compared with \gls{piv} experiments at selected cross-sections along the center-line. The results from R1, R2, and R3 are indicated by {solid black, dashed black and solid red} lines respectively. Cross-section planes X$_1$ to X$_8$ are represented from top left to bottom right. Note the different vertical scales between the two rows.}
 \end{figure*}

\paragraph{Viscous shear stress}
The magnitude of the mean viscous shear stress component ${\sigma}_{xy}$ along the same cross-sections is illustrated in \Cref{fig:Re500ViscousShearStress}. Overall, numerical viscous shear stresses are in good agreement with the values reported from the \gls{piv} measurements. Peak stresses between the central jet and the recirculation zones show a decreasing trend in downstream direction after entering the sudden expansion. The viscous shear stresses are consistently zero along the center-line.

\subsection{Transitional flow, Re $= 3500$}

\begin{figure}[htbp]
 	\centering
 	\includegraphics[width=0.42\textwidth]{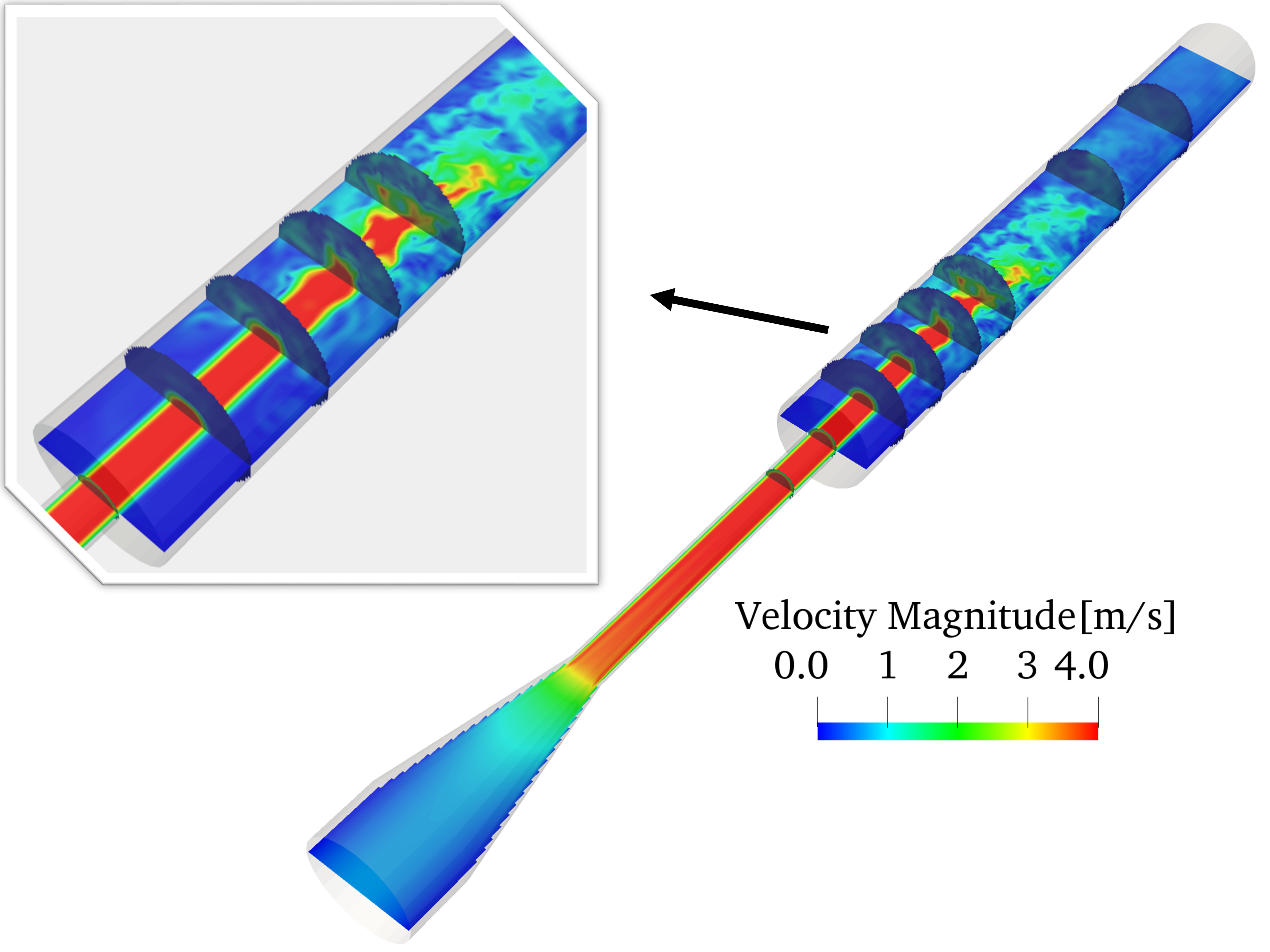}
 	\caption{\label{fig:Re3500FlowField} Results for Re=3500. Snapshots of instantaneous velocity field as obtained from R3 simulation at time $t = 4.2$\,s. The top-left inlay is a zoom over the sudden expansion.}
\end{figure}

At Re $= 3500$ (inlet Reynolds number of 1169), the flow regime has been observed to be transitional with a jet breakdown downstream of the sudden expansion region observed both in the \gls{piv} experiments and numerical simulations (see \Cref{fig:Re3500FlowField}). 

\begin{figure}[htbp]
 \centering
 \includegraphics{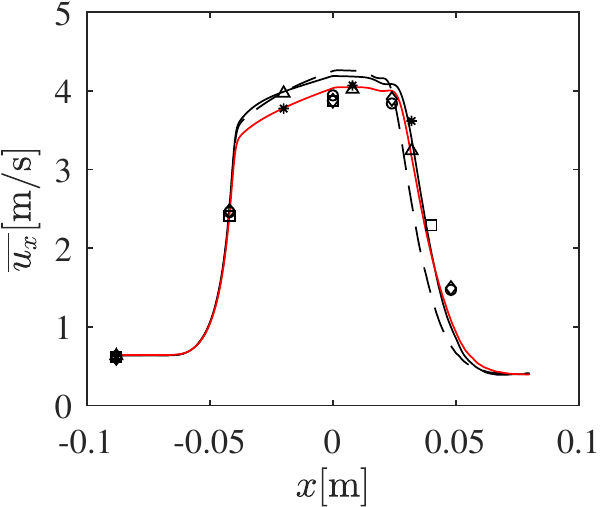}
 \caption{Results for Re $= 3500$. Comparison of mean axial velocity between \gls{lbm} results and \gls{piv} experiments along the center-line of the nozzle. The results from R1, R2 and R3 are indicated by {solid black, dashed black and solid red} lines respectively.}
 \label{Re3500_pressure_Velocity}
 \end{figure}

\begin{figure*}[htbp]
 \centering
 \includegraphics[width=0.8\textwidth]{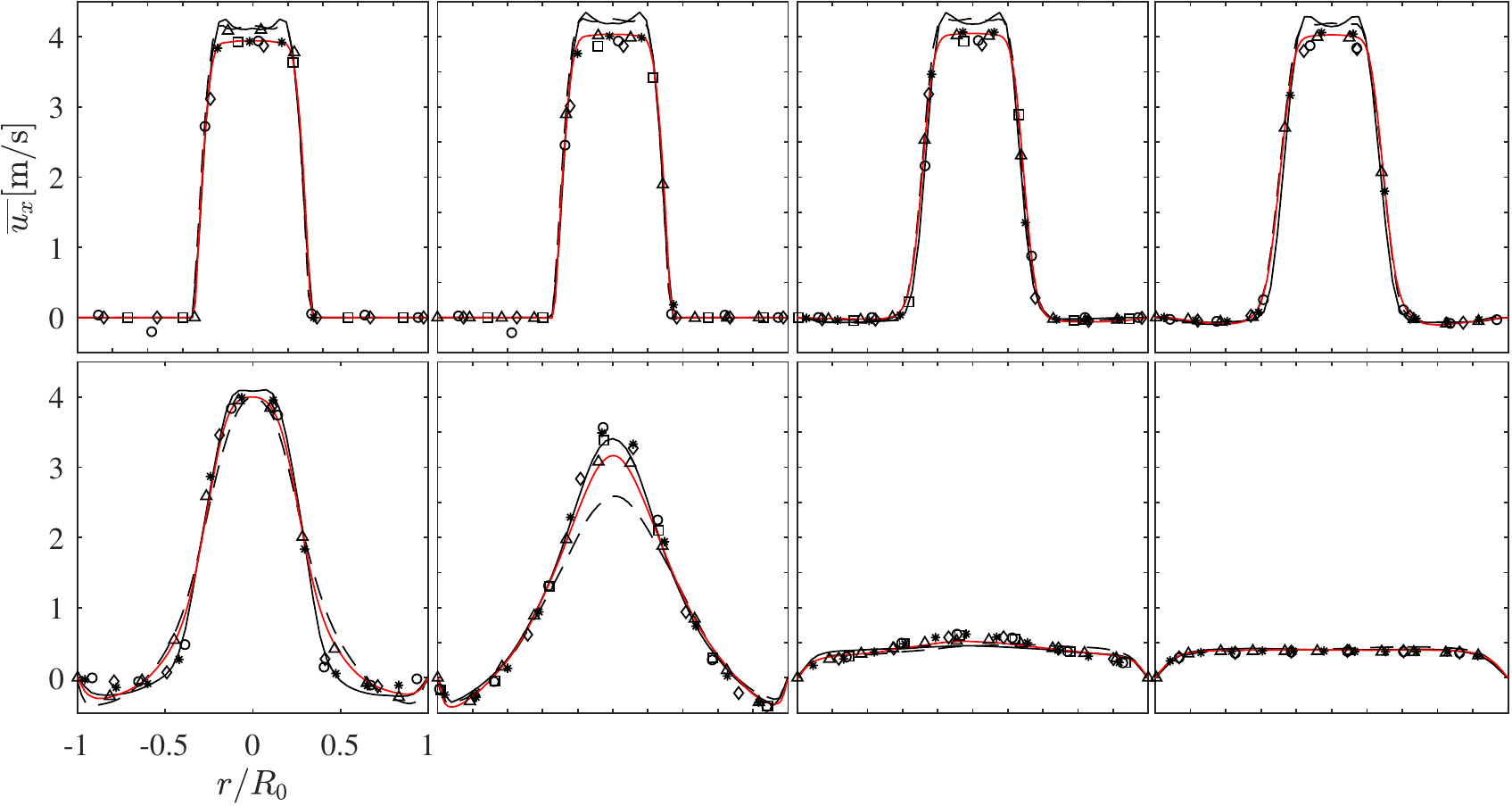}
 \caption{\label{fig:Re3500Axial_Vel_RadialDirection} Results for Re = 3500. Time-averaged axial velocity profiles compared with \gls{piv} experiments at selected cross-sections. The radial distance is normalized by the radius of the nozzle, $R_0$. The results from R1, R2 and R3 are indicated by {solid black, dashed black and solid red} lines respectively. Cross-section planes X$_1$ to X$_8$ are represented from top left to bottom right.}
 \end{figure*}

\paragraph{Axial velocity}Looking at \Cref{Re3500_pressure_Velocity}, the axial velocity reaches a peak of 4 m/s on the finest grid (R3) at $x = 0.02$ m, but then decreases afterwards rapidly to $0.5$ m/s at $x = 0.06$ m, which indicates jet break-down. The velocity distributions obtained from all three simulations match perfectly up to the smooth contraction zone. In the throat, small discrepancies (of the order of 3\% relative differences) are observed in the predicted maximum velocities. Overall, the time-averaged axial velocities along the center-line predicted by \gls{lbm} simulations match the \gls{piv} data very well, particularly so for resolution R3. The differences observed between R1, R2 and R3 are in the same range as the variations found in the measurements.
\paragraph{Radial profiles of axial velocity on sampling planes}
\Cref{fig:Re3500Axial_Vel_RadialDirection} shows axial velocity profiles along the chosen cross-section planes. In contrast to the parabolic profiles found at Re = 500, the axial velocity profiles are now plug-like in the throat section, as expected.  Starting at $X_4$, negative axial velocities appear close to the walls indicating a large, recirculating toroidal zone. Further downstream, the profiles slowly grow into a parabolic distribution with a decreasing peak value, down to $0.5$ m/s at the late stages, i.e. the last plane $X_8$. Visible differences between different grid resolutions are observed most notably at $\rm{X_6}$, where the highest levels of turbulence (and hence smallest structures) are present, as will be discussed later. The results with the finest mesh R3 ({red} lines in \Cref{fig:Re3500Axial_Vel_RadialDirection}) lead everywhere to an excellent agreement with measurement data.
\begin{figure*}[bp]
 \centering
 \includegraphics[width=0.8\textwidth]{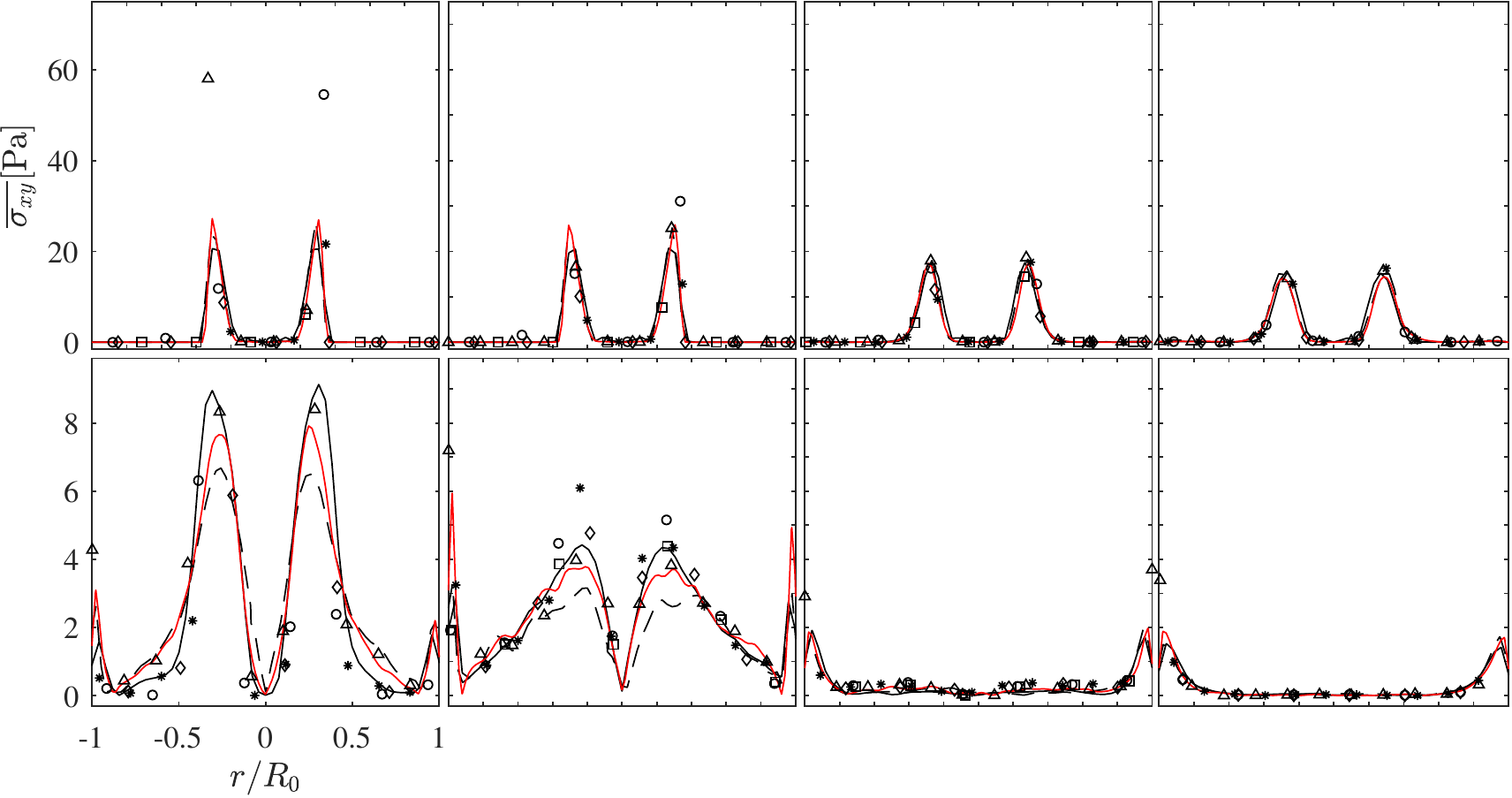}
 \caption{\label{fig:Re3500ViscousShearStress} Results for Re = 3500. Time-averaged viscous shear stress compared with \gls{piv} experiments at selected cross-sections. The results from R1, R2 and R3  are indicated by {solid black, dashed black and solid red} lines respectively. Cross-section planes X$_1$ to X$_8$ are represented from top left to bottom right. Note the different vertical scales between the two rows.}
 \end{figure*}
 \begin{figure*}[tp]
 	\centering
 	\includegraphics[width=0.8\textwidth]{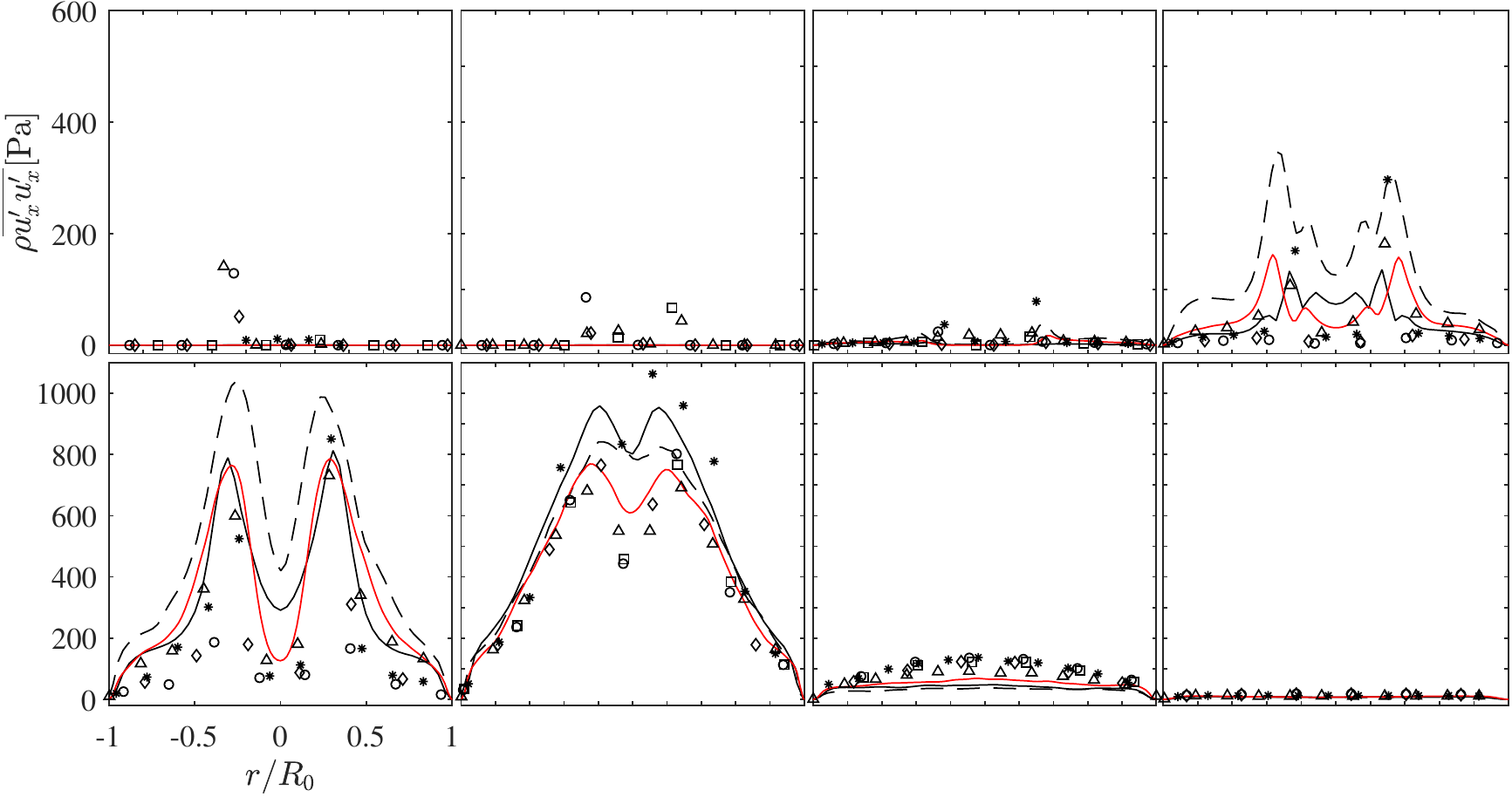}
 	\caption{\label{fig:Re3500ReynoldsNormalStress_radial} Results for Re $=3500$. Time-averaged normal Reynolds stress in axial direction compared with \gls{piv} experiments at selected cross-sections. The results from R1, R2 and R3 are indicated by {solid black, dashed black and solid red} lines respectively. Cross-section planes X$_1$ to X$_8$ are represented from top left to bottom right. Note the different vertical scales between the two rows.}
 \end{figure*}

  \begin{figure}[htbp]
 	\centering
 	\includegraphics{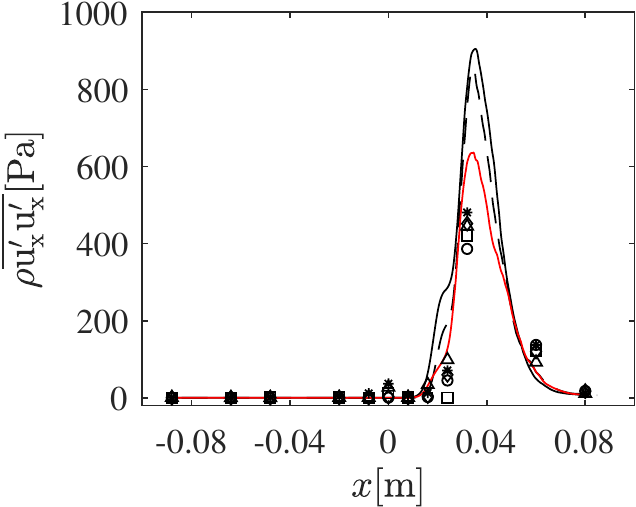}
 	\caption{\label{fig:Re3500ReynoldsNormalStress_axial} Results for Re$=3500$. Comparison between \gls{lbm} results and experimental data for normal Reynolds stress in axial direction along the center-line. The results from R1, R2 and R3  are indicated by {solid black, dashed black and solid red} lines respectively.}
 \end{figure}
Even strongly under-resolved simulations, like R1, lead to fully correct and stable qualitative predictions, and stay even in fair quantitative agreement with measurement data outside of the transitional region, demonstrating the robustness of the employed numerical solver.

\paragraph{Viscous shear stress}
 The magnitude of the mean viscous shear stress component ${\sigma}_{xy}$ is shown in \Cref{fig:Re3500ViscousShearStress}. The peak of the viscous shear stress is found at around 25$\,$Pa in the throat section. The stress decreases gradually from plane $\rm{X_2}$ to the end of the nozzle. The peak is shifted toward the shear layer between the jet and the recirculation zone during jet breakdown. The shear stress remains zero in the boundary layers of the recirculation region at planes $\rm{X_3}$ and $\rm{X_4}$, while it starts increasing there at plane $\rm{X_5}$, before progressively going back to zero further downstream (relaminarization).
 
\paragraph{Reynolds stress}
 \Cref{fig:Re3500ReynoldsNormalStress_axial,fig:Re3500ReynoldsNormalStress_radial} show the normal Reynolds normal in axial direction at selected cross-sections and along the center-line, respectively. Ahead of the sudden expansion section, the Reynolds stress is nearly zero, indicating laminar conditions.

  It experiences a sharp increase around plane $\rm{X_4}$, reaching maximum values in planes $\rm{X_5}$ and $\rm{X_6}$. The peak of the normal Reynolds stress is found to be around 800\,Pa for the finest resolution at $\rm{X_6}$, where strong flow instabilities can be observed from \Cref{fig:Re3500FlowField}. Further downstream (cross-section planes X$_7$ \& X$_8$) the turbulent stress goes back to zero, indicating again relaminarization of the flow.
  While noticeable differences are observed among the three different resolutions (especially around the jet breakdown region), they are observed to converge toward experimental data, as seen in \Cref{fig:Re3500ReynoldsNormalStress_axial}. The numerical results from the finest mesh R3 fall within the \gls{piv} data and follow closely the corresponding trends, leading to an accurate description of jet breakdown.
  It must also be kept in mind that an accurate measurement of the Reynolds stress tensor based on \gls{piv} remains extremely challenging due to the needed resolution in space and time. As such, the experimental data used for the present comparison is certainly associated with a non-negligible level of experimental uncertainty, already visible in the widely different results obtained by the three experimental groups involved in the study.

\subsection{Mildly turbulent flow, Re $= 6500$}
\begin{figure}[h]
 	\centering
 	\includegraphics[width=0.42\textwidth]{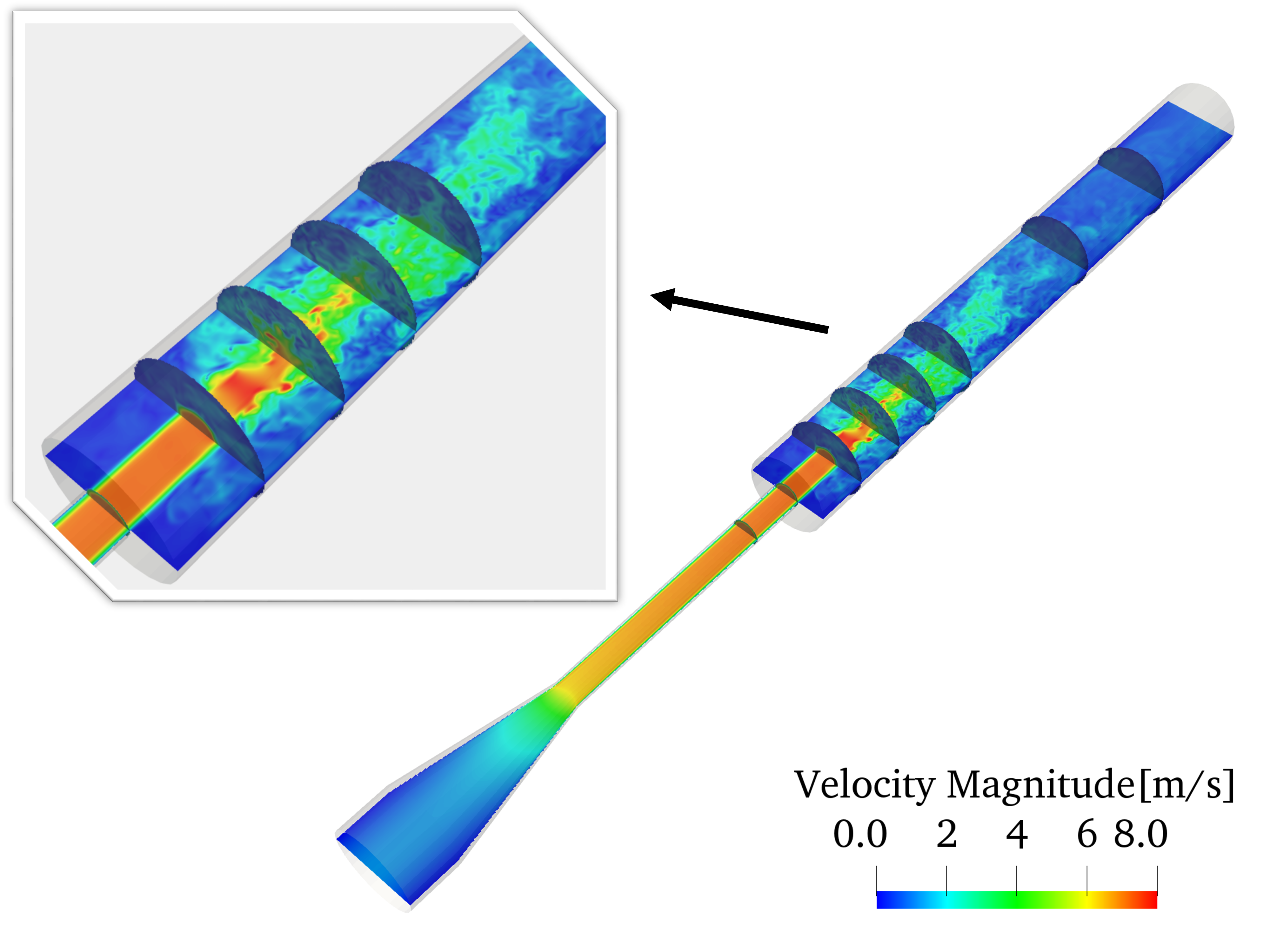}
 	\caption{Results for Re = 6500. Instantaneous axial velocity field for mesh R1 at time $t = 11.925$ s.}
 	\label{fig:Re6500FlowField_Resolution}
\end{figure}

At Re $= 6500$ (inlet Reynolds number of 2171), the flow regime is observed to be locally turbulent downstream of the sudden expansion region in the \gls{piv} and in the numerical simulations (see \Cref{fig:Re6500FlowField_Resolution}). 
\paragraph{Axial velocity}

\begin{figure}[hbtp]
 \centering
 \includegraphics{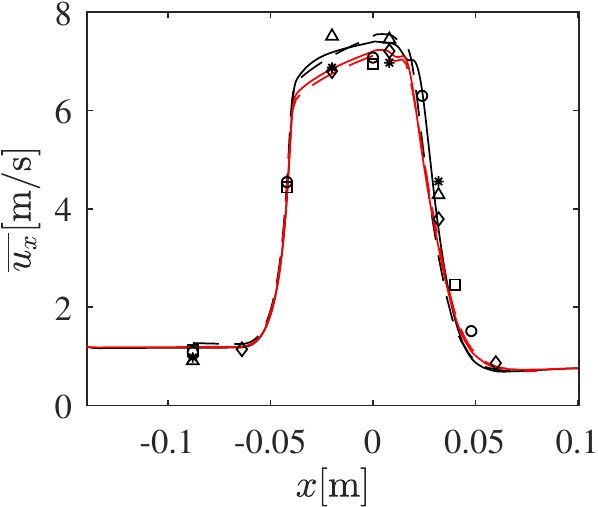}
 \caption{Results for Re = 6500. Comparison between \gls{lbm} results and \gls{piv} experiments for mean axial velocity along the center-line of the nozzle. The results from {solid black, dashed black, solid red and dashed red} lines respectively. 
 }
 \label{Re6500fig_pressureAndVelocity}
\end{figure}

The profiles of time-averaged axial velocity for the \gls{piv} data-sets and for the {four \hbox{\gls{lbm}} meshes ({R1}, {R2}, {R3} and {R4})} are shown along the center-line in \Cref{Re6500fig_pressureAndVelocity}. As in \Cref{Re3500_pressure_Velocity} the \gls{lbm} axial velocity profiles show very good agreement with \gls{piv} experimental data-sets. The results with {mesh resolution R3 and R4} fall within the experimental symbols. It is worth noting that the jet from the coarsest mesh, R1, starts breakdown slightly later than at higher resolution. The axial velocity increases to a peak value around $7.5$ m/s at $x = 0.01$ m, before decreasing very sharply to less than $0.5$ m/s. This abrupt decrease marks the onset of jet breakdown. All {four} resolutions are able to accurately predict the axial velocity distribution along the center-line throughout the \gls{fda} nozzle.

\begin{figure*}[hbtp]
 \centering
 \includegraphics[width=0.8\textwidth]{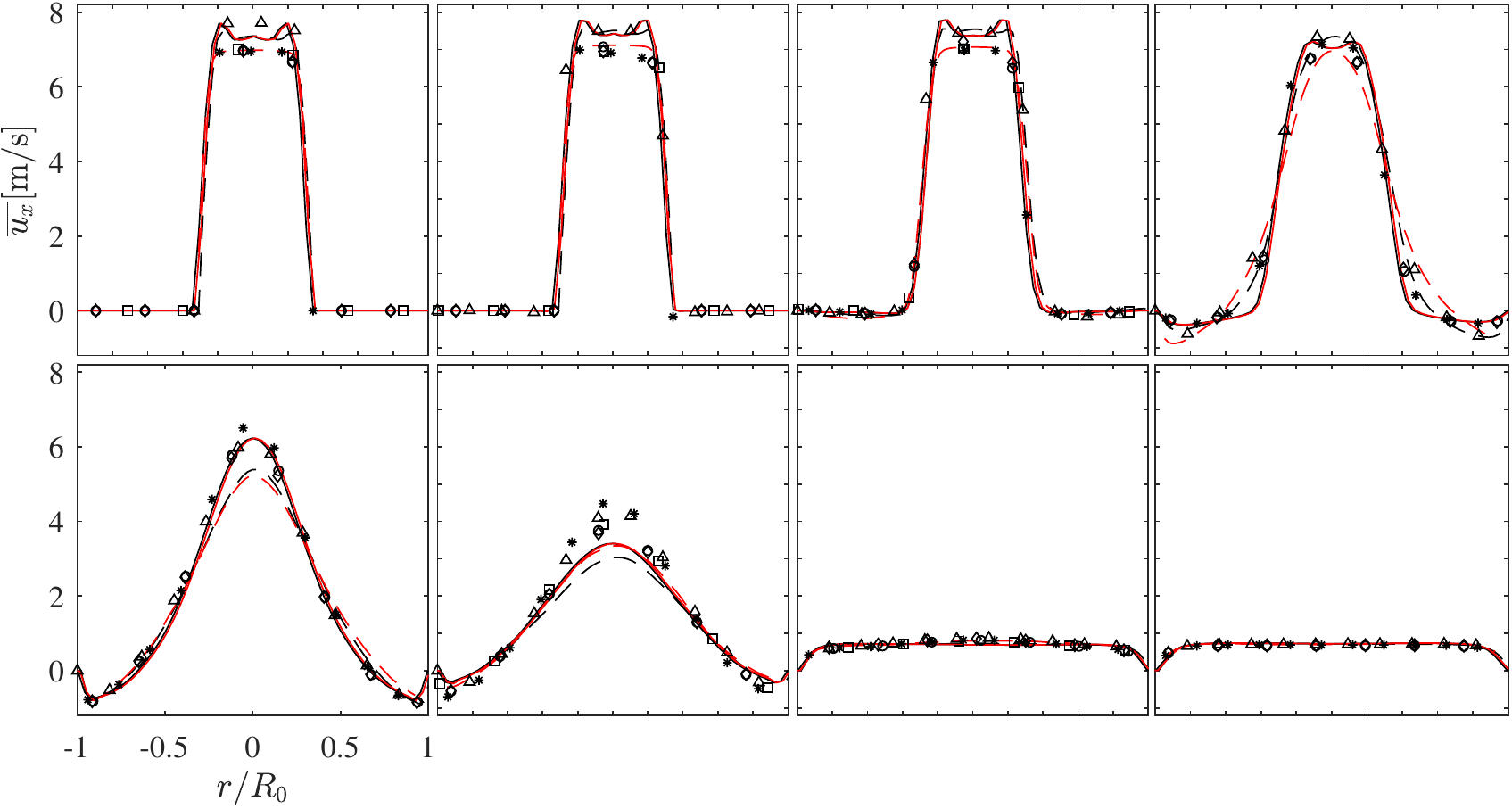}
 \caption{\label{fig:Re6500Axial_Vel_RadialDirection}Results for Re = 6500. Time-averaged axial velocity compared with \gls{piv} experiments at the chosen cross-sections along the center-line. The radial distance is normalized by the radius of the nozzle, $\rm{R_0}$. The results from R1, R2, R3 and R4 are indicated by {solid black, dashed black, solid red and dashed red} lines respectively. Cross-section planes X$_1$ to X$_8$ are represented from top left to bottom right.
 }
 \end{figure*}

\paragraph{Radial profiles of axial velocity along sampling planes}
The time-averaged axial velocity profiles are illustrated at the selected cross-sections along the flow direction in ~\Cref{fig:Re6500Axial_Vel_RadialDirection}. It can be seen from the first two plots that the flow remains at the same peak velocity in the throat section; the numerical results from the {four} resolutions match well with the \gls{piv} experiments.
 Although the flow enters the expansion section at location $\rm{X_3}$, the peak axial velocity is still the same as in the previous locations, in good agreement with \gls{piv} data.
 It is worth noting that a small negative velocity appears at cross-section $\rm{X_4}$ near the boundary layers, while the peak axial velocity decreases only slightly. On the other hand,
 the peak axial velocity decreases drastically from section $\rm{X_5}$ to $\rm{X_{6}}$ in association with jet breakdown.
 At later stages, in the cross-sections $\rm{X_{7}}$ and $\rm{X_{8}}$, it can be observed that the peak values are relatively small (around $0.75$ m/s). The numerical predictions match very well with the \gls{piv} data in this region where the jet collapses almost completely and the flow reattaches. Overall, the \gls{lbm} results are in very good agreement with the experiments concerning jet breakdown. {The mesh} R3 captures correctly the axial velocity profile and the jet breakdown process throughout the nozzle.

\begin{figure*}[htbp]
 \centering
 \includegraphics[width=0.8\textwidth]{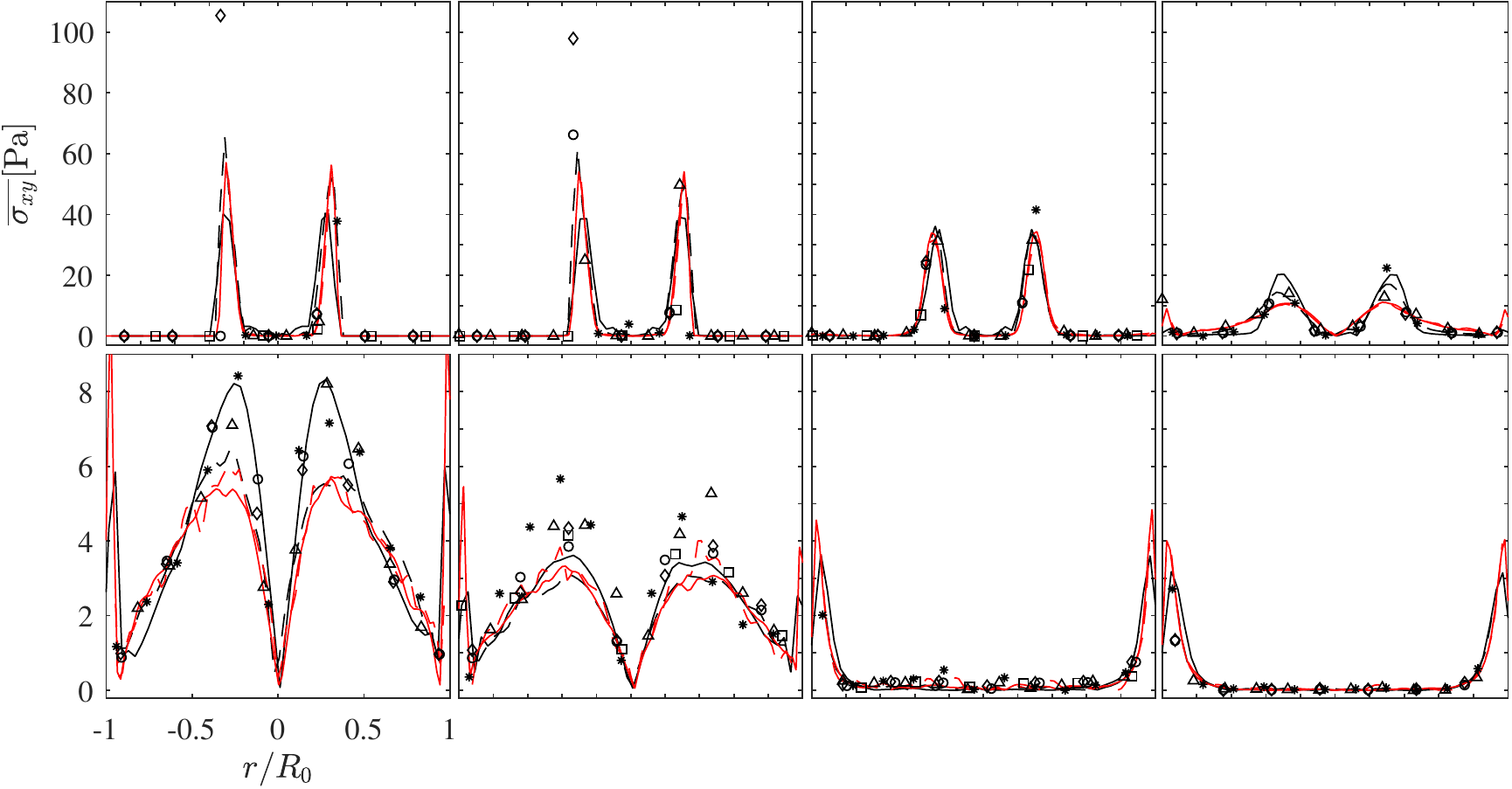}
 \caption{\label{fig:Re6500ViscousShearStress.pdf}Results for Re = 6500. Time-averaged viscous shear stress compared with \gls{piv} experiments at selected cross-sections along the center-line. The results from R1, R2, R3 and R4 are indicated by {solid black, dashed black, solid red and dashed red} lines respectively. Cross-section planes X$_1$ to X$_8$ are represented from top left to bottom right. Note the different vertical scales between the two rows.
 }
 \end{figure*}

\paragraph{Viscous shear stress}
 The magnitude of the time-averaged viscous shear stress component ${\sigma}_{xy}$ is shown at selected cross-sections in \Cref{fig:Re6500ViscousShearStress.pdf}. 
 It can be noted that the viscous shear stress reaches a peak around 50 Pa in the throat section, the largest velocity gradients being found as expected in the boundary layer. In the sudden expansion section, the peak of the viscous shear stress is predicted to be less than 8 Pa and is found in the shear layers between the central, high-velocity jet and the recirculating fluid. The peak of the viscous shear stress decreases further downstream.  Later on, the effect of viscous shear stress can be fully neglected due to the uniform and low velocity found in the nozzle. The largest (but low) shear stress is found close to the boundary layer in cross-sections $\rm{X_{7}}$ and $\rm{X_{8}}$, which can be attributed to the jet reattachment to the wall. As expected, the shear stress is zero along the center-line. The height of the boundary layer is estimated to be around half of the throat diameter in both experimental and numerical results.

 Overall, viscous shear can be neglected as the peak viscous stress is very small throughout the nozzle. Such a value should not be able to damage red blood cells. Depending on the studies, a shear stress exceeding about 100 Pa is necessary to observe any hemolysis \cite{yu}.
 
\begin{figure*}[htbp]
 	\centering
 	\includegraphics[width=0.8\textwidth]{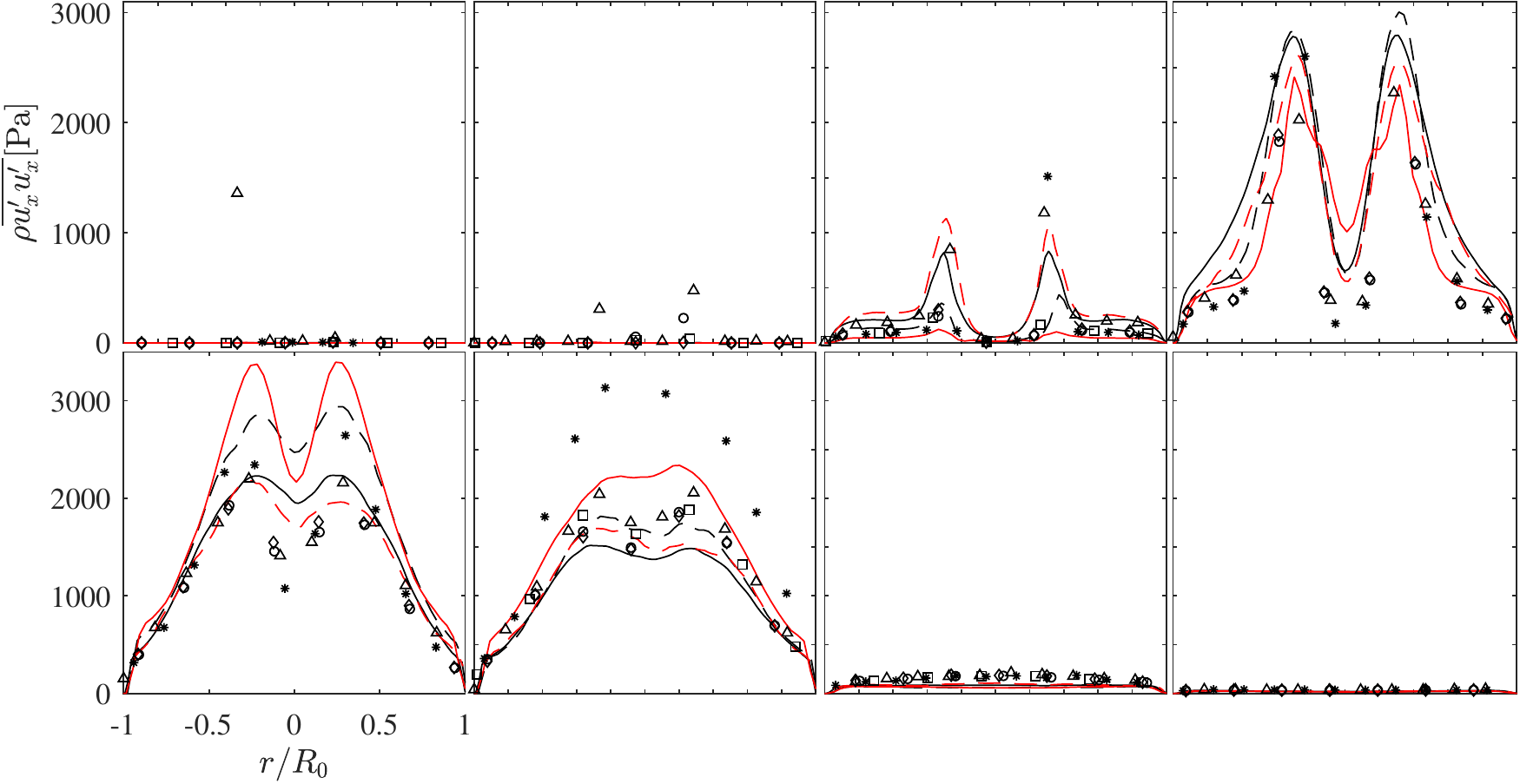}
 	\caption{\label{fig:Re6500ReynoldsNormalStress_radial} Results for Re=6500. Comparison between \gls{lbm} results and experimental data for axial Reynolds normal stress at different cross-sections. The results from R1, R2, R3 and R4 are indicated by {solid black, dashed black, solid red and dashed red} lines respectively. Cross-section planes X$_1$ to X$_8$ are represented from top left to bottom right.}
 \end{figure*}

\begin{figure}[htbp]
 	\centering
 	\includegraphics{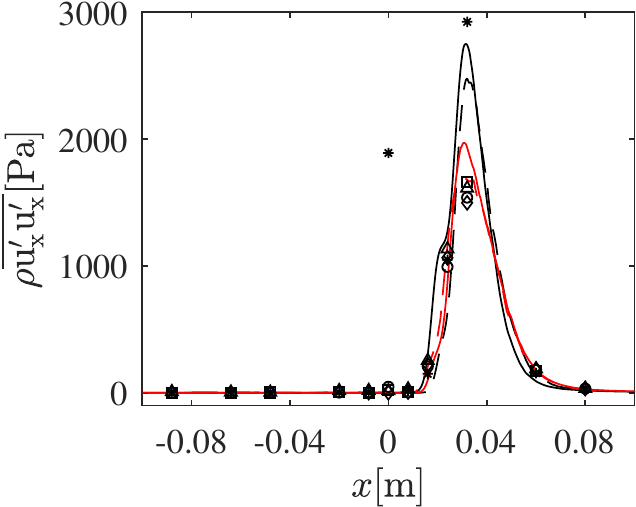}
 	\caption{\label{fig:Re6500ReynoldsNormalStress_axial} Results for Re=6500. Comparison between \gls{lbm} results and experimental data for axial Reynolds normal stress along the center-line. The results from R1, R2, R3 and R4 are indicated by {solid black, dashed black, solid red and dashed red} lines respectively.}
 \end{figure}

\paragraph{Reynolds stress}
 \Cref{fig:Re6500ReynoldsNormalStress_axial,fig:Re6500ReynoldsNormalStress_radial} illustrate the normal Reynolds stress for selected cross-planes and along the center-line, respectively. It can be observed that the results are in good agreement with the experimental data throughout the nozzle, keeping in mind the large variations also found in the experiments. In the throat section (planes $\rm{X_1}$\& $\rm{X_2}$), the numerical results are close to zero, as in most measurements (while few experimental sets show local peaks up to 1500 Pa; the experimental results vary very widely there). After flowing through the sudden expansion region, the Reynolds stress increases largely until the late stage of jet breakdown. The peak of Reynolds stress appears in this region in the shear layers between the central high-speed jet and the surrounding recirculating fluid, reaching around 3000 Pa between cross-sections $\rm{X_{5}}$ and $\rm{X_{6}}$. Further downstream, the Reynolds stress starts decreasing sharply towards zero. The Reynolds stress is negligible in the throat as well as at the end of the nozzle.

\section{Discussion\label{sec:discussion}} 
In this section, further information is provided concerning agreement and/or discrepancies between numerical results and \gls{piv} measurements in this study. 
\paragraph{Laminar flow} The in-house numerical tool ALBORZ successfully predicts the nature of the laminar flow. The jet induced after leaving the throat does not break down in the sudden expansion section. The numerical simulations accurately capture the fields of velocity and viscous shear stress.

\paragraph{Transitional flow} Both numerical simulations and experiments show a transitional regime with a jet breakdown in the sudden expansion section at this Reynolds number of 3500. The location of the jet breakdown is accurately predicted by the simulations, around $x = 0.02$ m, which is later than for Re = 6500. The effect of the viscous shear stress would be negligible regarding hemolysis, due to its low peak value. Concerning the Reynolds stress, used to quantify the strength of turbulent velocity fluctuations and to assess indirectly the potential damage to blood cells. The obtained values are much higher and would be sufficient to trigger hemolysis. However, as explained in Yu et al.~\cite{yu}, there is still no clear consensus regarding critical values and unsteady effects must be taken into account to clearly assess the impact of stress on hemolysis.

\paragraph{Mildly turbulent flow}  
The turbulent flow exhibits an earlier onset of jet break-down, already around $x = 0.01$ m.  Ahead of the sudden expansion section, the value of the Reynolds stresses are small, showing that there is no notable flow disturbance produced before and within the throat, as expected. Some discrepancies are found between numerical results and \gls{piv} measurements regarding Reynolds normal stress for cross-section planes $\rm{X_5}$ and $\rm{X_{6}}$. However, the differences are below the variations observed among the different experimental data-sets, the orders of magnitude are well predicted, and the shapes of the profiles coincide. It is therefore impossible to decide if the disagreements come from the numerical model or from experimental limitations.
\section{Conclusions}\label{sec:conclusion}
Lattice Boltzmann simulations were performed on the \gls{fda}'s benchmark nozzle, mimicking a simplified blood-contacting medical device, to assess the capability of the in-house solver ALBORZ in modelling such flows under laminar, transitional, and mildly turbulent conditions. For this purpose, Reynolds numbers Re = 500, 3500, and 6500 have been considered, for which experimental data are available.
Velocity, viscous shear stress, and turbulent Reynolds stress were compared to the inter-laboratory \gls{piv} experiments documented in the literature~\cite{hariharan2011multilaboratory}. Overall, numerical results are in very good agreement with the experimental data, demonstrating that the solver is capable of predicting accurately the underlying flow physics. This applies even for under-resolved simulations (grid R1), allowing therefore very fast estimates, since results on grid R1 are typically obtained almost 20 times faster than on grid R3. When a very high accuracy is required, by increasing the resolution, results converge toward experimental profiles within the corresponding uncertainty. This shows that the solver can be used in a stable manner either 1) for a fast estimate of the flow conditions with a fair accuracy, or 2) to get very accurate flow predictions at the price of a much longer computational time.
To our knowledge, this work is the first \gls{lbm} study of the \gls{fda}'s benchmark nozzle considering simultaneously laminar, transitional, and mildly turbulent flow conditions, providing comprehensive quantitative comparisons for the flow fields.
\section*{Acknowledgements}
 F.H. would like to acknowledge the financial support of China Scholarship Council (grant number 201908080236). S.A.H. acknowledges funding by the Deutsche Forschungsgemeinschaft (DFG, German Research Foundation) in TRR 287 (Project-ID 422037413). P.B. acknowledges the funding by the Federal Ministry of Education and Research in Germany within the Forschungscampus STIMULATE (grant number 13GW0473A) and the German Research Foundation (grant number BE 6230/6 -1). The authors also gratefully acknowledge the Gauss centre for providing computation time under grants {"pn29du" and} "pn73ta" on the GCS supercomputer superMUC-NG at Leibniz supercomputing centre, Munich, Germany.
\section*{Declaration of competing interest}
 None.
\appendix
\section{Central Hermite moments collision operator}
{In the present work the continuous particle-speed space is discretized via a third-order Gauss-Hermite quadrature. The third-order quadrature leads to stencils of the form DdQ$3^d$, where $d$ is the number of physical dimensions, i.e. $d\in\{1,2,3\}$. Furthermore the third-order quadrature yields the following weights in 1-D:}
\begin{equation}
    w_{i,\alpha}=\begin{cases} 2/3 & \text{for } c_{i,\alpha} = 0,\\
    1/6 & \text{for } c_{i,\alpha} = \delta r/\delta t,
    \end{cases}
\end{equation}
{which in turn for the 3-D stencil is merely the tensorial product of the 1-D weights:}
\begin{equation}
    w_i = \prod_{\alpha=x,y,z} w_{i,\alpha},
\end{equation}
{and $\theta_0=\delta r^2/3\delta t^2$. The Hermite polynomials appearing in Eq.~\hbox{\ref{eq:equilibrium}} are defined as:}
\begin{subequations}
	\begin{align}
	\mathcal{H}_0 &= 1, \\
	\mathcal{H}_\alpha &= c_{i,\alpha}, \\
	\mathcal{H}_{\alpha\beta} &= c_{i,\alpha} c_{i,\beta} - \delta_{\alpha\beta}, \\
	\mathcal{H}_{\alpha\beta\gamma} &= c_{i,\alpha} c_{i,\beta} c_{i,\gamma} - {\rm cyc}(\delta_{\alpha\beta} c_{i,\gamma}), \\
	\mathcal{H}_{\alpha\beta\gamma\zeta} &= c_{i,\alpha} c_{i,\beta} c_{i,\gamma} c_{i,\zeta} - {\rm cyc}\left( \delta_{\alpha\beta} c_{i,\gamma} c_{i,\zeta}\right) \nonumber\\ &+ {\rm cyc}\left( \delta_{\alpha\beta}\delta_{\gamma\zeta} \right),\\
	\mathcal{H}_{\alpha\beta\gamma\zeta\iota} &= c_{i,\alpha} c_{i,\beta} c_{i,\gamma} c_{i,\zeta} c_{i,\iota} - {\rm cyc} \left( \delta_{\zeta\iota} c_{i,\alpha} c_{i,\beta} c_{i,\gamma} \right) \nonumber\\ &+ {\rm cyc} \left(c_{i,\alpha}\delta_{\beta\gamma}\delta_{\zeta\iota}\right),\\
	\mathcal{H}_{\alpha\beta\gamma\zeta\iota\kappa} &= c_{i,\alpha} c_{i,\beta} c_{i,\gamma} c_{i,\zeta} c_{i,\iota} c_{i,\kappa} - {\rm cyc}\left(c_{i,\alpha} c_{i,\beta} c_{i,\gamma} c_{i,\zeta} \delta_{\iota\kappa} \right)\nonumber\\ & + {\rm cyc}\left(c_{i,\alpha} c_{i,\beta} \delta_{\gamma\zeta}\delta_{\iota\kappa}\right) - {\rm cyc}\left(\delta_{\alpha\beta}\delta_{\gamma\zeta}\delta_{\iota\kappa}\right),
	\end{align}
\end{subequations}
{with ${\rm cyc}$ a cyclic permutation and resulting equilibrium coefficients:}
	\begin{subequations}
		\begin{align}
		\bm{a}_0^{\rm eq} &= \rho, \\
		\bm{a}_\alpha^{\rm eq} &= \rho u_\alpha, \\
		\bm{a}_{\alpha\beta}^{\rm eq} &= \rho u_\alpha u_\beta, \\
		\bm{a}_{\alpha\beta\gamma}^{\rm eq} &= \rho u_\alpha u_\beta u_\gamma, \\
		\bm{a}_{\alpha\beta\gamma\zeta}^{\rm eq} &= \rho u_\alpha u_\beta u_\gamma u_\zeta,\\
		\bm{a}_{\alpha\beta\gamma\zeta\iota}^{\rm eq} &= \rho u_\alpha u_\beta u_\gamma u_\zeta u_\iota,\\
		\bm{a}_{\alpha\beta\gamma\zeta\iota\kappa}^{\rm eq} &= \rho u_\alpha u_\beta u_\gamma u_\zeta u_\iota u_\kappa.
		\end{align}
	\end{subequations}
{In the central moments space, assuming a D3Q27 stencil, the following equilibrium moments are recovered:}
\begin{equation}
    \bm{T} \bm{f}^{\rm eq} = \widetilde{\bm{a}}^{\rm eq},
\end{equation}
{where $\widetilde{\bm{a}}^{\rm eq}$ are the equilibrium Hermite coefficients in central moments space:}
\begin{subequations}
\begin{alignat}{4}
		\widetilde{\bm{a}}^{\rm eq}_0 &= \rho, \\
		\widetilde{\bm{a}}^{\rm eq}_n &= 0, \forall n\neq 0.
\end{alignat} \label{eq:Hermite_coeff_eq}
\end{subequations}
{It must be noted that all equilibrium moments of order $n>0$ become zero in central space only using the full Hermite expansion supported by the stencil. In the present work we use a sixth-order Hermite-expansion of the discrete equilibrium for all simulations.\\
Apart from the relaxation rates of second-order moments tied to the fluid viscosity, {i.e.} $a_{xx}$, $a_{yy}$, $a_{zz}$, $a_{xy}$, $a_{xz}$ and $a_{yz}$, all other relaxation rates are set to 1. A set of Matlab code are included as supplementary material to compute the transform and inverse tensors along with corresponding moments for the collision model used in the present work.\\
It is also interesting to note that the viscous stress tensor obtained as the second-order moment of the non-equilibrium part of the distribution function~\hbox{\cite{kruger2009shear}} can be readily obtained from Hermite coefficients:}
\begin{equation}
    \frac{1}{2}\left(\partial_\alpha u_\beta + \partial_\beta u_\alpha\right) = \frac{1}{2\rho \theta_0\bar{\tau}}\left(a_{\alpha\beta} - a^{\rm eq}_{\alpha\beta}\right),
\end{equation}
{or central Hermite coefficients:}
\begin{equation}
    \frac{1}{2}\left(\partial_\alpha u_\beta + \partial_\beta u_\alpha\right) = \frac{1}{2\rho \theta_0\bar{\tau}}\left(\widetilde{a}_{\alpha\beta}\right).
\end{equation}
\bibliography{references}
\end{document}